\documentclass{article}

\usepackage{graphicx}
\usepackage{amsmath}
\usepackage{amssymb}
\usepackage{latexsym}

\newtheorem{theorem}{Theorem}[section]
\newtheorem{sled}{Consequense}[section]

\newtheorem{statement}{Statement}[section]
\newtheorem{zam}{Remark}[section]
\newtheorem{lemma}{Lemma}[section]

\title{The properties of the distribution of Gaussian
packets on a spatial network.}
\author{V.\,L. Chernyshev, A.\,A. Tolchennikov}

\begin{document}
\maketitle

\begin{abstract}
The article deals with the description of the statistical behavior of
Gaussian packets on a metric graph. Semiclassical asymptotics of
solutions of the Cauchy problem for the Schr\"{o}dinger equation with
initial data concentrated in the neighborhood of one point on the
edge, generates a classical dynamical system on a graph. In a
situation where all times for packets to pass over edges ("edge travel times") are linearly
independent over the rational numbers, a description of the behavior
of such systems is related to the number-theoretic problem of counting
the number of lattice points in an expanding polyhedron. In this paper
we show that for a finite compact graph packets almost always are
distributed evenly. A formula for the leading coefficient of the
asymptotic behavior of the number of packets with an increasing time
is obtained. The article also discusses a situation where the times of
passage over the edges are not linearly independent over the
rationals.
\end{abstract}

\section{Introduction.}
In this paper we study semiclassical solutions of the Cauchy problem for a time-dependent Schr\"{o}dinger equation on a metric graph. This article is, in a s certain sense, a continuation of \cite{cts}, \cite{mian} where you can find necessary definitions and formulations (nevertheless, we quote all necessary terms here, in section \ref{notation}).  There (and in \cite{mz}) are also references to some articles and reviews related to the study of differential equations on metric graphs.

In \cite{mian} a formula for the asymptotics of the number of Gaussian packets was written for a star graph, and a question of finding the leading coefficient for the case of an arbitrary finite compact graph was raised. In this paper that problem is solved and a general formula for that coefficient is obtained (see Theorem \ref{th_koef}).
Moreover, in \cite{mian}  the uniformity of distribution of packets for the special case of two vertices connected by three edges was proven. In this paper we present the proof of uniformity of the asymptotic distribution of packets over an arbitrary finite compact graph for almost all edge travel times (see Theorem \ref{th_ravnom}).

In the second part (section \ref{neq}) we discuss a description of the asymptotic behavior of packets in the case of linearly dependent over $\mathbb{Q}$ edge travel times. In this situation there is no correspondence (see \cite{cts}, \cite{mian} for details)  between the number of packets and the number of nodes of an integer lattice that lie on some faces of an expanding simplex, but analysis of their number is still possible.  If rank of the system of edge travel times (lengthes) is equal to one then the number of packets grows only in a finite interval of time and stabilizes at a certain value, which depends on the lengths of cycles. Also we consider an example of a graph, whose rank of the system of lengthes equals two.

\subsection{Acknowledgment.}

V.\,L. Chernyshev thanks M.\,M. Skriganov, N.\,G. Moschevitin, P.\,B. Kurasov and O.\,V. Sobolev for useful discussions and attention to his work. Authors are grateful to  A.\,I. Shafarevich for constant attention to their work. The work is done with partial finance support of grants MK-943.2010.1, RFFI 10-07-00617-a and 09-07-00327-a, RNP 2.1.1/11818 and state contract 14.740.11.0794.

\subsection{Terms and Definitions.}
\label{notation}

Recall (see, e.g., \cite{pokkniga}, \cite{ku} and references therein), that a {\it metric graph} is a one-dimensional cell complex whose edges are parameterized curves. We denote a geometric graph by $\Gamma$, its edges by $\gamma_j$, its vertices by  $a_j$. A set of all edges adjacent to the vertex $a$ we denote by $ \Gamma(a)$.
We consider only finite metric graphs. Let $E$ and $V$ stand for the number of edges and the number of vertices respectively.

\par Let $Q$ be an arbitrary real valued continuous function on $ \Gamma $, smooth
on the edges.
{\it
Schr\"{o}dinger operator}
\begin{equation}
\widehat{H}\psi=-{h^2}\frac{d^2 \psi(x)}{d x^2} +Q(x)\psi(x)\label{shrodik}
\end{equation}
is defined on the set of functions from Sobolev spaces
 $\psi\in\oplus\sum\limits_{j}H^2(\gamma_j)$, satisfying the following
boundary conditions at the vertices:
\begin{enumerate}
\item
 function $\psi$ is continuous on $\Gamma$;
\item
\begin{equation}
\sum\limits_{\gamma_j\in\Gamma(a_m)}\alpha_j\frac{d \psi_j} {d x} (a_m)=0,
\label{trans_nashe}
\end{equation}
in all internal vertices (i.e., the vertices of degree
greater than one);
\item
$\psi(a_m)=0$ in all the external vertices, i.e. vertices of
degree one.
\end{enumerate}

Here $\alpha_j=1$ for each edge emerging from the vertex, and $\alpha_j=-1$ for each incoming edge.

These conditions are called {\it natural} (see \cite{kurasov}). They, in particular, ensure self-adjointness of the operator $\widehat{H}$.

{\it A time-dependent Schr\"{o}dinger equation on the graph} $\Gamma$ is an equation of the form

\begin{equation} \begin{array}{cc} \displaystyle
ih\frac{{\partial}\psi}{{\partial}t}=\widehat{H}\psi,
\label{shrodik_n} \end{array} \end{equation}

where a semiclassical parameter $h\to 0$.
We choose initial conditions that have the form of a narrow packet localized near the point $x_0$, which lies on the edge of the graph:

\begin{equation}
\psi(x,0)=h^{-1/4}K\exp\left(\frac{iS_0(x)}{h}\right).
\label{nach_dannye} \end{equation}

$$
S_0(x)=a(x-x_0)^2 + b(x-x_0) + c,
$$

where $b$ and $c$ are real constants, and $a$ and $K$ are complex. The imaginary part of $a$ is positive. Normalization factor
$h^{-1/4}$ is introduced to ensure that the initial function $\psi(x, 0)$ is of order one in the norm of $L^2(\Gamma)$. Due to the positivity of the imaginary part of $a$ the initial function is localized in a small neighborhood of $x_0$: $\psi(x, 0)=O(h^{\infty})$ with $| x-x_0 |\geq\delta>0$ ($\delta$ is independent of $h$). For simplicity we assume that there are no turning points (see, e.g., \cite{mf}, \cite{cwkb}) on the $\Gamma$ ; their presence can be accounted for in the standard way (see \cite{mf}).

Asymptotic solution of the Cauchy problem (\ref{shrodik_n})--(\ref{nach_dannye}) is described in \cite{mian}, the explicit formulas are given therein.

\begin{theorem}  (See \cite{mian}) Solution of the Cauchy problem \ref{shrodik_n} -- \ref{nach_dannye} for $ t \in [0, T] $ ($T$ does not depend on $h$), is given by the following formula

\begin{equation}
\begin{array}{lr} \displaystyle
\psi(x,t,h)=\sum_{j=1}^{N(t)}h^{-1/4}\varphi_j(t)e^{iS_j(x,t)/h}+O(\sqrt{h}),
\label{rostok}
\end{array}
\end{equation}

where the functions $\varphi_j(t), S_j(t)$ are explicitly expressed in terms of the solutions of two hamiltonian systems. \end{theorem}
Each term in the sum (\ref{rostok}) is localized in a small neighborhood of $X_j(t)$. Here we assume that all the terms that are localized in the same point $X_j(t)$ form one {\it Gaussian packet}. Later, under the $N(t)$ we would mean the number of such packets.

In \cite{mian} it is shown that in moments of penetration of the vertices of the graph a quantum packet is divided into $m$ packets ($m$ is a degree of the vertex): one is reflected and $m-1$ are ``diffused''. The amplitude of the packet is divided as $(m-2):2 $ (2 for the scattered and $(m-2)$ for the reflected packet).

We consider the asymptotical behavior of function $\psi(x,\,t\,h)$ as $t{\rightarrow}\infty$. Namely, we will see how the number of Gaussian packets $N(t)$ changes in time.
Note that this problem differs from the task of describing the asymptotic solution of the Schr\"{o}dinger equation at $t{\rightarrow}\infty$, as the error estimation is valid only for finite times. From a physical point view, this means that we are considering big $t$, but much smaller than $1/h$.

Let $t_j$ stand for $j$-th edge travel time (this is an analog of length).

\begin{lemma} Travel time of any edge of the graph depends only on the initial data and is the same for any Gaussian packet on each fixed edge.
\end{lemma}

\textbf{Proof.} Let us take an edge with index $j$. By construction a solution on each edge (using the method of Maslov complex germ), we
find that  $P^2+Q(x)=E_j$ holds (by the energy conservation law for
Hamiltonian systems), where the value of $E_j$ is determined by the
initial conditions. Writing $P$ as $\frac{d X}{dt}$, we obtain for edge travel time on the edge an explicit expression
$t_j=\int\limits_a^b\frac{dx}{\sqrt{E_j-Q(x)}}$, where the integral is
taken over the edge. The value of $E_j$ is the same for all edges, as
the potential $Q(x)$ is assumed to be continuous
function in the vertices, and $P$ can only change sign (as demonstrated by a construction of the
solution). So the formula $E_j=P^2+Q(x)$ determines the same value
$E_j=E$ for all packets. At the initial moment of time
$E=\left(\frac{\partial S_0} {\partial x} \right)^2 + Q(x_0)$. Values
of $t_j$ may be different, since the restriction of the potential
$Q(x)$ on the edges may be different.

{\bf Definition.} Let us consider the number of packets coming out of a fixed vertex to a fixed edge. The leading coefficient of the asymptotics of this number is called a radiation coefficient.

Correctness of this definition will be shown in the proof of the theorem \ref{th_ravnom}.

\section{The case of linearly independent over $\mathbb{Q}$ edge travel times.}

\subsection{Main results.}

\begin{theorem}[About uniformity of distribution]

Consider a finite connected graph $\Gamma$. Suppose that for any vertex its degree is not equal to two. Suppose that there are no turning points on edges. Consider an edge $e$. Let $F_e(t)$ be a ratio of $N_e(t)$ (number of packets on the edge $e$) to $N(t)$ (total number of packets).
Then for almost all incommensurable (i.e. linearly independent over $\mathbb{Q}$) numbers $t_1,\dots t_E$, a ratio of $F_e(t)$ to the length of $e$ tends to a constant $\left(\sum\limits_{j=1}^{E}t_j\right)^{-1}$ as time increases.

\label{th_ravnom}
\end{theorem}

\begin{zam}
It means that the distribution of number of packets tends to a uniform distribution as time increases.
\end{zam}

\textbf{Proof.} Let us choose on any edge with travel time $t_j$ a segment $dg$ with travel time $\tau$. Let us find $N_\tau(t)/N(t)$. We know (see \cite{mian}) that $N(t)=Ct^{E-1}+o(t^{E-1})$.  Let us find $N_\tau(t)$. Since the number of packets changes only in vertices and there are no turning points, then:

\begin{equation}
N_\tau(t)=N_{{\rightarrow}d}(t)-N_{{\rightarrow}d}(t-\tau)+N_{{\rightarrow}g}(t)-N_{{\rightarrow}g}(t-\tau).
\label{Ntau}
\end{equation}

Here $N_{{\rightarrow}d}(t)$ stands for the number of packets which arrived at the edge from point $d$.

It is clear that the number of packets arrived at point $d$ at time $t$ equals the number of packets, which came out from the nearest vertex $a$ at time $t-T_1$. Here $T_1$ is a travel time from $a$ to $d$. By $N_{a{\rightarrow}d}(t)$ we denote the number of packets, which came from $a$ to $d$.

We have to know asymptotics of the number of packets that come out of a vertex $a$. Packets can come out of the vertex only at times that are linear combinations (with nonnegative integer coefficients) of edge travel times.

The number of release moments (when at least one packet comes out of the vertex $a$) is described by the number of set $\{n_j\}$ satisfying inequations of a kind:

\begin{equation}
n_1t_{l_1}+\ldots+n_mt_{l_m}{\leq}t, \label{ineq}
\end{equation}
where $t_j$ is a travel time of the $j$-th edge.

Since the leading part of asymptotics of the number of packets is defined by the volume of a simplex defined by (\ref{ineq}), events with maximal numbers of summands happen more often. In other words, packets arrived at our vertex should visit all edges. I.e.

\begin{equation}
N_{a{\rightarrow}d}(t)=R^at^{E}+o(t^{E}).
\label{Ra}
\end{equation}

For almost all $t_1,\dots t_E$ the estimation can be improved (see \cite{Skriganov}). There exists $K^a$ such that $N_{a{\rightarrow}d}(t)=R^at^{E}+K^at^{E-1}+o(t^{E-1})$.
Let us show that $R^a$, which is called {\it a radiation coefficient}, does not depend on the choice of a vertex. Consider vertices $a$ and $b$.  There exists a path connecting $a$ and $b$. Let $\delta$ be its travel time. Any packet coming out from $a$ to $d$ over time that does not exceed $2\delta$ generates at least one packet that come out from $b$ to $d'$. This is correct for packets coming out from $b$. We obtain inequations: $N_{a{\rightarrow}d}(t+2\delta){\geq}N_{b{\rightarrow}d'}(t)$ and $N_{b{\rightarrow}d'}(t+2\delta){\geq}N_{a{\rightarrow}d}(t)$.
We know that $N_{a{\rightarrow}d}(t)=R^at^E+o(T^E)$, $N_{b{\rightarrow}d'}(t)=R^bt^E+o(T^E)$. Thus $R^at^E+o(t^E)=R^bt^E$. Hence $R^a=R^b$.

Let us modify the expression for $N_\tau(t)$

$N_\tau(t)=R(t-T_1)^{E}+K^a(t-T_1)^{E-1}-R(t-T_1-\tau)^{E}-K^a(t-T_1-\tau)^{E-1}+R(t-T_2)^{E}+K^b(t-T_2)^{E-1}-R(t-T_2-\tau)^{E}-K^b(t-T_1-\tau)^{E-1}+o(t^{E-1})=2ER{\tau}t^{E-1}+o(t^{E-1})$.

Thus we obtain

\begin{equation}
\frac{N_\tau(t)}{N(t)}{\rightarrow}\frac{2ER}{C}\tau.
\label{otv}
\end{equation}

It remains to show that a coefficient near $\tau$ has the required form.

We consequentially take edges as $dg$ and then summarize obtained expressions:

\begin{equation}
1=\sum\limits_{j=1}^{E}\frac{N_{t_j}(t)}{N(t)}{\rightarrow}\frac{2ER}{C}\sum\limits_{j=1}^{E}t_j.
\label{C_C1}
\end{equation}

Hence,

\begin{equation}
C=2ER\sum\limits_{j=1}^{E}t_j.
\label{C_R_1}
\end{equation}

The proof is completed.

\begin{zam}

For any two vertices numbers $K^a$ and $K^b$ are related with inequation $|K^b-K^a|{\leq}E{\delta}R$. Here ${\delta}$ is a travel time for any path from $a$ to $b$.
\end{zam}

In the proof we obtain the following statement:

\begin{sled}[Relation between coefficients $C$ and $R$]

The leading coefficient for the number of packets $C$ and the radiation coefficient $R$, for almost all edge travel times, are related in the following manner:

\begin{equation}
C=2ER\sum\limits_{j=1}^{E}t_j.
\label{C_R}
\end{equation}
\end{sled}

\begin{theorem}[About the leading coefficient of the number of packets]
Consider a finite connected compact graph $\Gamma$. Suppose that there are no vertices of degree two. Suppose that there are no turning points on edges. Then for almost all incommensurable numbers $t_1,\dots t_E$ the  leading coefficient has the following form:

\begin{equation}
C=\frac{1}{2^{V-2}(E-1)!}\frac{\sum\limits_{j=1}^{E}t_j}{\prod\limits_{j=1}^{E}t_j}.
\label{form_C}
\end{equation}

\label{th_koef}
\end{theorem}

\textbf{Proof} is based on (\ref{form_C}) and the following lemma.

\begin{lemma} Let us consider a finite connected graph with incommensurable edge travel times  $t_i (i=1 \ldots E)$ and a number of independent cycles $\beta$.
Let B be an arbitrary vertex. Then for almost all edge travel times the number of packets arriving at B at time T asymptotically equals

$$
R(T) \sim \frac{2^{\beta}}{2^E\, E!\, {\prod\limits_{j=1}^{E}t_j}} T^E.
$$
\end{lemma}

\begin{zam}

This is equivalent to
$$
R=\frac{1}{2^{V-1}E!}\frac{1}{\prod\limits_{j=1}^{E}t_j}.
$$
\end{zam}

\textbf{Proof.} Is is sufficient to consider paths of packets that traveled upon all edges. Only those paths give us the leading coefficient. Let A be an initial vertex. For each such path we can construct a ``code'' i.e. a sequence of coefficients of a corresponding chain with coefficients in $\mathbb{Z}_2$. It is clear that the code does not change under a path homotopy. Let us find the number of all possible codes.
Consider cross connections, i.e. edges that are not in the spanning tree. Parity of passages on cross connections defines a path's homotopy class. All chain coefficients are defined by coefficients on cross connections. Thus the number of possible codes equals $2^\beta$. Now for every code $(c_1, \ldots, c_E), c_i \in \{ 0, 1 \}$ we associate times
$$
\left\{ \sum_{i=1}^E t_i (c_i + 2 n_i) | n_i \in \mathbb{N} \cup \{ 0 \} \right\}.
$$
At every such time (they are different) at least one packet arrives at the vertex $B$.
The number of such times that are less than $T$ asymptotically equals to
$$
\frac{T^E}{2^E \, E! \, t_1 \cdots t_E}.
$$
Finally we summarize this over all possible codes. The proof of the lemma is finished.
At the end we apply Euler's relation $\beta=E-V+1$ (see, for example, \cite{chris}).

\section{The case of linearly dependent travel times.}
\label{neq}

In this section we assume that travel times are linearly dependent over $\mathbb{Q}$.
It means that there is no one-to-one correspondence (described in \cite{cts}, \cite{mian}) between the number of packets and the number of integer lattice points in an expanding simplex.

For the simplest example consider a star graph with three edges of the same length. The number of packets reaches three and does not increase. While the number of integer lattice points grows with time.

\begin{statement} Consider a finite graph with edge travel times  $t_1, = n_1t_0   \ldots,t_E = n_E t_0$, where $n_i \in \mathbb{N}$ and
GCD$(n_1,\ldots,n_E) = 1$.\\

Then starting from a certain time, the number of packets becomes constant.\\

1) If there exists a cycle with travel time $2 k t_0, k \in \mathbb{N}$ then
$$
N(T) = 2 \sum_{i=1}^{E} n_i.
$$
2) Otherwise
$$
N(T) = \sum_{i=1}^{E} n_i.
$$
\end{statement}

We omit the proof.

\begin{statement} Consider a star graph with three edges $e_1, e_2, e_3$ with travel times
$t_1 = n t_0, t_2 = m t_0, t_3$, where $n \in \mathbb{N}, m \in \mathbb{N}$ (GCD(n,m) = 1), and $t_3$ is such that rank $\{t_1, t_2, t_3\}$
over $\mathbb{Q}$ equals 2. Then the number of packets asymptotically equals
$$
N(T) = \frac{T}{2}\left( \frac{ m + n }{t_3} + \frac{1}{t_0} \right) + o(T).
$$
\end{statement}

\textbf{Proof.} The number of packets changes only at times of a kind $2((n \alpha + m \beta)t_0 + t_3 \gamma )$,
where $\alpha, \beta, \gamma$. Note that this representation is not unique. If we can present a time moment with
$\alpha \ge 1, \beta \ge 1, \gamma \ge 1$, then at that moment $N(T)$ does not change (since packets arrive at a vertex from all three edges)

\begin{zam} Number $n \alpha$ can be represented as $n\alpha' + m \beta'$ ($\beta' \ne 0$) if and only if $\alpha \ge m$.
\end{zam}

Hence to define asymptotics of $N(T)$ it is sufficient to consider the following times:
\\
1) $\gamma \ne 0, \beta = 0, 1 \le \alpha < m$. At those times $N(T)$ increases by 1. The number of such times equals the number of pairs $(\alpha, \gamma)$ such that
$2(\alpha t_0 + \gamma t_3) < T$. The number asymptotically equals: $(m-1)\frac{T}{2t_3}$.
\\
2) $\gamma \ne 0, \alpha = 0, 1 \le \beta < n$. Similarly the number of pairs asymptotically equals $(n-1)\frac{T}{2t_3}$. \\
3) $\alpha = 0, \beta = 0$. The number of such times that are less than $T$ increases as
$\frac{T}{2t_3}$. Here the number of packets increases by 2.
\\
4) $\gamma = 0$. Let us show that linear combinations $\alpha n + \beta m$ contain all natural numbers that are greater than a certain fixed number.

\begin{lemma} If $(m,n) = 1$ then there exists $M$ such that any $N > M$ can be presented as $N = \alpha n + \beta m$,
where $\beta \ge 1, \alpha \ge 1 $.
\end{lemma}

\textbf{Proof.} Numbers $m, 2m, \ldots, nm$ give all residues modulo $n$:
$$
\{ \beta m \}_{\beta = 1}^{n} = \{\beta_i m = i + n k_i | i \in [0,n-1], k_i \in \mathbb{Z}_{+} \}.
$$
Let $k_0 = \max k_i$. Choose any $N > nk_0$. Then $N = i + nk \ (k > k_0, 0 \le i < n)$ and
$N - \beta_i m = n (k - k_i)$. Thus any number that is greater than $n k_0$ can be represented as $\alpha n + \beta m$,
where $1 \le \beta \le n, \alpha \ge 1 $.

Asymptotics of the number of such times coincides with asymptotics of numbers $2kt_0 (k \in \mathbb{N})$ that are less than $T$. Since at such times 2 packets arrive from $e_1$ and $e_2$, then $N(T)$ increases by 1.


\begin{thebibliography}{99}

\bibitem{Skriganov} Skriganov M.\,M.  {Ergodic theory on ${\rm
 SL}(n)$, Diophantine
approximations and anomalies in the lattice point problem.}
// Invent. Math. 132, no. 1, 1998. p.\,1--72.

\bibitem{cts}
Tolchennikov A.A., Chernyshev, V.L., Shafarevich A.I.  Asymptotic
properties of the classical dynamical systems in quantum problems on
singular spaces / / Nonlinear Dynamics, 2010, v.6, No.3, p.623-638 (in
Russian).

\bibitem{mian}
Chernyshev V. L., Time-dependent Schr\"{o}dinger equation: statistics
of the distribution of Gaussian packets on a metric graph // Proc.
Steklov Inst. Math., 270 (2010), p.246--262.

\bibitem{mz}
Chernyshev, V.L., Shafarevich A.I. Semiclassical spectrum of the
Schr\"{o}dinger operator on a geometric graph // Mathematical Notes,
Volume 82, Numbers 3-4, 2007, p.542-554.

\bibitem{chris}
Christofides N., Graph Theory -- An Algorithmic Approach, Academic
Press, London, 1975. 415 p.


\bibitem{cwkb}
Maslov V.\,P.,
{The Complex WKB Method for Nonlinear Equations 1: Linear
Theory.} Birkh\"{a}user-Verlag, Basel, 1994.

\bibitem{mf} Maslov V.\,P.,, Fedorjuk M.\,V., {Semi-classical approximation in
quantum mechanics}, D.Reidel, Dordrecht, Holland, 1981.

\bibitem{pokkniga} Pokornyi Yu.\,V., Penkin O.\,M., Pryadiev V.\,L., Borovskikh A.\,V., Lazarev K.\,P., Shabrov S.\,A., {Differntial equations on
geometrical graphs.} //(in Russian), Moscow: Fizmatlit, 2004.

\bibitem{ku} Kuchment P. {Quantum graphs: an introduction and a brief survey} -- ArXiv 0802.3442v1 [math-ph] 23 Feb 2008.

\bibitem{kurasov} Kurasov P., Nowaczyk M., {Inverse spectral problem for quantum graphs.}  // J. Phys. A: Math. Gen. 38, 2005. p.\,4901--4915.

\end{thebibliography}
\end{document}